# Optical Black-hole Analog Created by Topological Phase Transition with a Long-lived Horizon

Meng Kang[1,2], Huaqing Huang[2], Shunping Zhang[1], Hongxing Xu[1,3*] and Feng Liu[2†]

[1]*School of Physics and Technology, Center for Nanoscience and Nanotechnology, and Key Laboratory of Artificial Micro- and Nano-structures of Ministry of Education, Wuhan University, Wuhan 430072, China*

[2]*Department of Materials Science and Engineering, University of Utah, Salt Lake City, Utah 84112, USA*

[3]*The Institute for Advanced Studies, Wuhan University, Wuhan 430072, China*

**E-mail**: [*]hxxu@whu.edu.cn, [†]fliu@eng.utah.edu

**Abstract:** Hawking radiation, a manifestation of quantum field theory in curved spacetime, has stimulated extensive theoretical and experimental studies of various black-hole (BH) analogs. However, an undisputed confirmation of Hawking radiation remains elusive. One challenge is BH analog structures with long-lived horizons are difficult to achieve. Here, we theoretically demonstrate a new type of optical BH analog based on light cone evolution associated with topological phase transition of Dirac cones. The transition from a type-II to type-I Dirac/Weyl cone creates an analogous curved spacetime that crosses a type-III Dirac/Weyl cone, which affords a stationary configuration of long-lived event horizon. Photons tunneling through the horizon emit a spectrum of Hawking radiation. As an example, we design a laboratory version in an inhomogeneous two-dimensional graphyne-like topological photonic lattice with a Hawking temperature of 0.14 $m$K. Understanding Hawking-like radiation in this unique topological BH is not only of fundamental interest in its own right but may also provide new hints to gravitational physics.

Hawking radiation [1] is derived based on general relativity, quantum mechanics and thermodynamics, and therefore has been considered as a critical testbed for the underlying connection between those physical domains. However Hawking temperature for a black hole (BH) with one solar mass is even lower than that of cosmic microwave background, which makes it difficult to be detected directly. Unruh published a seminal theoretical work to simulate Hawking radiation from a sonic black hole analog [2]. The propagation of a sound wave in a convergent trans-sonic fluid flow in the long-wavelength limit is mathematically equivalent to the behavior of a scalar wave in the spacetime of BH. This has inspired researchers to pursue a variety of laboratory artificial BHs, including water waves [3-5], Bose-Einstein condensates [6-8], superfluid helium [9], Fermi-degenerate flow [10], ion rings [11], polaritons [12, 13], magnons [14] and optics [15-20]. However, although some promising experimental verification of Hawking radiation has been performed in analog systems [4, 5, 7, 8, 13, 18-20], the claims from these observations are still under dispute by the community [21-28].

One challenge is BH analog structures with long-lived event horizons are difficult to achieve experimentally [28], which makes the interpretation of experimental data ambiguous. Recent attempts in nonlinear optical BHs based on quantum fluids of light have offered hope to generate stationary BH configurations with a single black horizon and high Hawking temperature [13, 29]. On the other hand, topological photonics [30-32] emerges as a rapidly growing field to explore artificial topological phases resembling condensed matter physics, which makes it possible for a new type of BH analog. Here, we propose theoretically a design of optical BH analogs via topological transitions between Dirac/Weyl bosons. It uniquely combines the following four features, which are only partially covered by previously proposed analogs: (i) Dirac/Weyl points are topologically protected and immune to perturbation; (ii) Dirac/Weyl cones correspond directly to light cones in curved spacetime; (iii) the propagation of light follows quantum mechanics; (iv) the horizon arising from a stationary "spacetime" configuration is long-lived.

It has been shown recently that topological phase transition from type-II to type-I Dirac/Weyl cones in inhomogeneous systems enables a unique platform for the simulation of BH and Hawking radiation, as the same form of transition of light cones occurs near the horizon of BH in general relativity [33, 34]. In such inhomogeneous schemes, the type-II region generates an artificial BH where particles are trapped. The transition interface between the type-I and type-II Dirac cone (DC), i.e., the type-III region, forms the horizon. The fermionic version of such BH analogs have been designed for Dirac/Weyl semimetals lately [33-37], where quantum mechanics enables electrons to escape from the BH before reaching the equilibrium state. However, experimental realization of such inhomogeneous topological electronic systems can be rather challenging. For example, while the topological flat bands in an electronic Kagome lattice [38-40] remains to be demonstrated, its optical counterparts have been already made in artificial optical Kagome lattices [41, 42]. Also, this electronic analog of Hawking radiation is a short-lived transient process as electron-hole pairs regain the equilibrium state the moment the tilted Dirac cones are created. These considerations have motivated us to propose an optical version in a graphyne-like photonic lattice (PL) [43], which has been recently shown to form optical type-I, type-II and type-III Dirac points (DPs), with a tunable hopping term [44].

We first discuss our proposed analog of BH via a Painlevé-Gullstrand metric [45-48] form of the Hamiltonian near the DP and the associated quantum tunneling mechanics. Then we design a laboratory analog using a platform of an inhomogeneous graphyne-like PL. By gradually tuning the hopping term via a chain of waveguides [44], DC is tilted from type-II to type-I to generate a BH spacetime including an event horizon (type-III). When eigenmodes around DC beyond the horizon are excited by an external light souce, quantum tunneling enables photons trapped in the type-II region to escape into the type-I region, and hence simulates Hawking radiation. Our *ab initio* calculations show that the three types of DCs are readily achievable by a gradual decrease of refractive index of the waveguide chain. Finally we exhibit a BH analog with a long-lived horizon and a detectable Hawking temperature about 0.14 *m*K.

The effective Hamiltonian around the DPs in two-dimension can be obtained from the minimal low-energy Weyl Hamiltonian [36]:

$$H(k) = c_x(k_x - k_x^{DP})\sigma_x + c_y(k_y - k_y^{DP})\sigma_y + v_x(k_x - k_x^{DP})\sigma_0 \ , \qquad (1)$$

where $\sigma_x = \begin{bmatrix} 0 & 1 \\ 1 & 0 \end{bmatrix}$ and $\sigma_y = \begin{bmatrix} 0 & -i \\ i & 0 \end{bmatrix}$ are the first two Pauli matrixes, $\sigma_0$ is the unit matrix, $k_x^{DP}$ and $k_y^{DP}$ depend on the location of DP, $c_x$ and $c_y$ represent the tilt angle along the $k_x$ and $k_y$ direction in the momentum space, respectively, and only the tilted DC along $k_x$ determined by $v_x$ is considered. DC is tuned by $v_x$ to be type-I for $|v_x/c_x|<1$, type-II for $|v_x/c_x|>1$ and type-III for $|v_x/c_x|=1$ [FIG. 1(a-c)]. The corresponding band structure is $\beta_\pm(k_x,k_y) = v_x(k_x - k_x^{DP}) \pm \sqrt{c_x^2(k_x - k_x^{DP})^2 + c_y^2(k_y - k_y^{DP})^2}$, where the sign + (-) stands for the upper (lower) band.

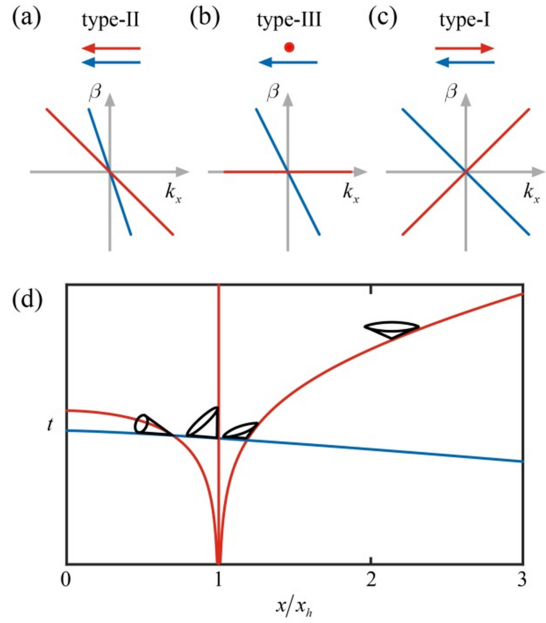

FIG. 1. Illustration of band diagrams close to the DP for (a) type-II, (b) type-III and (c) type-I DC. The direction of group velocity is shown above for counter-propagating (red arrowed line) and co-propagating (blue) photon modes. (d) Light trajectory in the created BH spacetime via the inhomogeneous PL (we assume $v_x = -c_x\sqrt{x_h/x}$ here). Light cones for counter-propagating (red) and co-propagating (blue) modes in type-II region ($x < x_h$), type-III region ($x = x_h$, the horizon) and type-I region ($x > x_h$) have the same features as an analogous particle–antiparticle pair in the astronomical BH.

To describe the effective spacetime, the Hamiltonian is rewritten as

$$H(k) = e_k^j(k_j - k_j^{DP})\sigma_k + e_0^j(k_j - k_j^{DP})\sigma_0, \qquad (2)$$

where Einstein summation convention is used, and the parameter $e_k^j = \delta_k^j c_j + \delta_k^0 v_j$ are the emergent tetrad fields in 2+1 dimension. According to general relativity, the line element takes the form $ds^2 = g_{\mu\nu} dx_\mu dx_\nu$ and the effective covariant metric $g_{\mu\nu}$ is defined by its inverse $g^{\mu\nu} = \eta^{ab} e_a^\mu e_b^\nu$ with $\eta^{ab} = diag(-1,1,1)$. As a result we have the line element in the Painlevé-Gullstrand coordinate [45-48]

$$ds^2 = -(1-\frac{v_x^2}{c_x^2})dt^2 - \frac{2v_x}{c_x^2}dtdx + \frac{1}{c_x^2}dx^2 + \frac{1}{c_y^2}dy^2 \ . \tag{3}$$

Here $c_{x(y)}$ is the light velocity and $v_x$ is the drag velocity of the effective fluid flow along the $x$ direction. When $dy = 0$ and $v_x(x) = -c_x\sqrt{x_h/x}$, $ds^2$ becomes the line element of BH in 1+1 dimensional spacetime, being stationary and nonsingular at the horizon at $x_h$.

The light trajectory [FIG. 1(d)] in the 1+1 dimensional spacetime rests on the light cone ($ds^2 = 0$, $dy = 0$) with two slopes $dt/dx = 1/(v_x \pm c_x)$, where + (-) represents the counter-propagating (co-propagating) light. The counter-propagating photons outside the horizon (type-I DC) will escape from the BH to emit a thermal Hawking radiation. When photons reach the horizon (type-III DC), their velocity becomes zero and will never enter the BH in the laboratory frame. Beyond the horizon (type-II DC), the dragging velocity exceeds the group velocity, which makes photons to be trapped into the artificial BH and impossible to escape. Hence this inhomogeneous PL with the designed DC transitioning gradually from type-II to type-I establishes a viable analog of BH to detect Hawking radiation.

Hawking radiation can be treated as a semiclassical quantum tunneling process in the vicinity of the horizon [49-51]. Here we describe Hawking radiation emitted from the inhomogeneous PL in the perspective of optical tunneling to derive Hawking temperature and the radiant spectrum. The horizon forms a potential barrier for the quantum tunneling. In the Wentzel-Kramers-Brillouin approximation, the tunneling probability is $\Gamma \sim \exp(-2S/\hbar)$ with $S = \text{Im} \int \hbar k_x(x)dx = \hbar\beta \, \text{Im} \int \frac{1}{x-x_h}dx \Big/ \frac{dv_x}{dx}\Big|_{x_h}$ being the imaginary part of the action for the trajectory forbidden in classical theory. This induces a Hawking temperature

$$T_H = \frac{\hbar}{2\pi k_B} \frac{dv_x}{dx}\bigg|_{x_h}, \quad (4)$$

which is equivalently a result of surface gravity $E_{gh} = dv_x/c_x dx\big|_{x_h}$ at the horizon. Here $k_B$ is Boltzmann constant. Photons quantum tunnel mechanically through the horizon to emit a Planck spectrum.

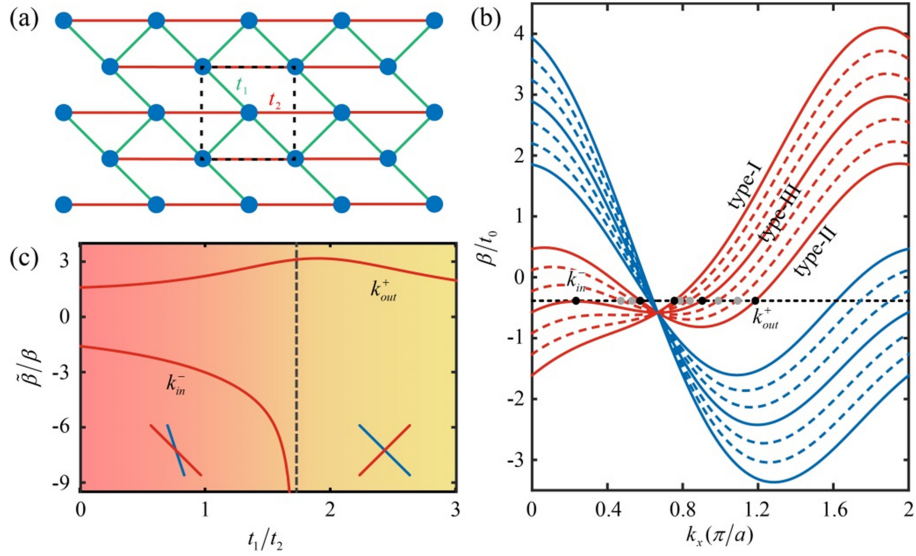

FIG. 2. (a) Illustration of the graphyne-like PL, which consists of a centered-square lattice with inter- and intra-hopping ($t_1$, $t_2$). (b) Evolution of calculated dispersion relation from type-II to type-I DC in lab-frame. Considering counter-propagating modes, an outgoing mode $k_{out}^+$ with positive norm and an incoming mode $k_{in}^-$ with negative norm constitute a Hawking pair. (c) Doppler shift in co-moving frame for outgoing and incoming modes. With the increase of hopping strength $t_1/t_2$, DC is gradually tuned from type-II to type-I. The dashed line represents the event horizon.

Next we demonstrate a scheme of the optical BH analog in an inhomogeneous PL based on a graphyne-like lattice [44], as shown in FIG. 2(a). Under the tight-binding approximation, the evolution of light diffraction in PL is described as

$$i\partial_z \psi_{mn}(z) = \sum_{m',n'} t_{mnm'n'} \psi_{m'n'}(z) , \quad (5)$$

where the hopping term $t_{mnm'n'}$ is the coupling strength between the two adjacent waveguides with the index ($m$, $n$) and ($m'$, $n'$), respectively, and $\psi_{mn}$ is the amplitude of the $m$, $n$th waveguide. The corresponding Hamiltonian becomes $H(\mathbf{k}) = \begin{pmatrix} h'(\mathbf{k}) & h^*(\mathbf{k}) \\ h(\mathbf{k}) & h'(\mathbf{k}) \end{pmatrix}$, with $h'(\mathbf{k}) = 2t_2 \cos k_x$, $h(\mathbf{k}) = t_1\{1 + \exp(-ik_y) + \exp[i(k_x - k_y)]\}$. The position of DP depends on $|h(\mathbf{k})| = 0$. Compared with Eq. (1), the angle of DC is related to hopping, $c_x = \partial \mathrm{Re}\{h(\mathbf{k})\}/\partial k_x$, $c_y = \partial \mathrm{Im}\{h(\mathbf{k})\}/\partial k_y$ and $v_x = \partial h'(\mathbf{k})/\partial k_x$, which means that only the intra-hopping term $t_2$ determines the tilt of DCs along $k_x$. It has already been made aware that two neighboring waveguides can couple with each other via an interactive waveguide chain. Consequently, the tuning among the three types of DC becomes flexible by perturbing the refractive index of the waveguide chain to adjust the intra-hopping $t_2$.

Hawking radiation can sometimes be confused because frequency is beyond the Planck scale [52-54] close to the horizon, where the physics is unknown, as shown in FIG. S1 (see supplymentary materials). In our analog system, however, the high frequency is cut off by the nonlinear dispersion relation in the group velocity $c_x(k_x)$, so that the trans-Planckian physics is clear. Hawking pairs constituted by two possible waves are attained for low (quasi)frequency, as shown in FIG.2(b). The outgoing mode $k_{out}^+$ has both positive group velocity and norm $\tilde{\beta}(p_x) = |c_x k_x|$ in co-moving frame, while the incoming mode $k_{in}^-$ has negative group velocity and norm $\tilde{\beta}(p_x) = -|c_x k_x|$. The positive and negative norm are associated with creation and annihilation operators, respectively. Their amplitudes α and γ are given by the Bugoliubov coefficients as

$$|\alpha/\gamma|^2 = \exp(\hbar\beta/k_B T_H). \qquad (6)$$

The negative norm mode stimulates the amplification of positive mode, which is an analog of Hawking radiation. The corresponding Doppler shift of the two modes in co-moving frame, $\tilde{\beta}(x) = \beta/[1 + v_x(x)/c_x]$, is shown in FIG. 2(c), where $\beta$ is conserved (quasi)frequency in lab-frame. The (quasi)frequency of outgoing mode becomes finite at the horizon due to dispersion relation.

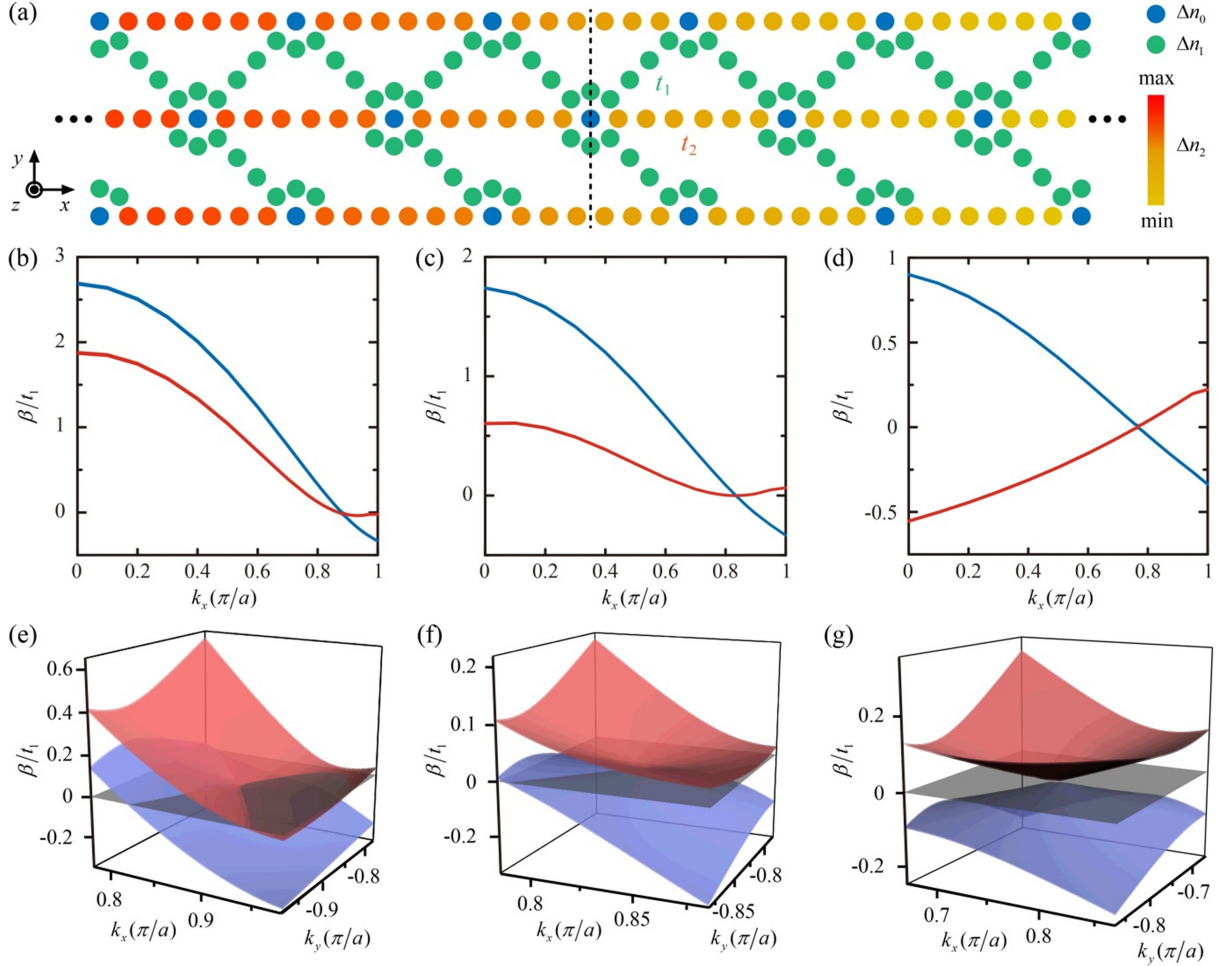

FIG. 3. (a) Two-dimensional sketch of the inhomogeneous graphyne-like PL. It contains a centered-square lattice of waveguides (blue dots) with a period $a = 30\sqrt{2}$ μm, whose inter- and intra-hopping ($t_1$, $t_2$) is built by a chain of waveguides (green, red and orange dots, respectively). The intra-hopping $t_2$ is tuned by the waveguide chains with a decreasing gradient of refractive index along the $x$ direction. Additional waveguides are used to maintain local symmetry. The interface (dashed line) between type-II and type-I region constitutes the horizon. Numerical simulation results of band diagram for (b) type-II DC with a conical-like isofrequency surface (the diameter of each waveguide d = 3.768 μm, the refractive index $\Delta n_0$ = $\Delta n_1$ = 0.0016, $\Delta n_2 = 2\Delta n_0$), (c) type-III DC (d = 4 μm, $\Delta n_2 = 1.69\Delta n_0$) and (d) tilted type-I DC with a point-like isofrequency surface (d = 4.479 μm, $\Delta n_2 = \Delta n_0$). The corresponding enlarged band structures in the first Brillouin zone are shown in (e-g), respectively.

The tunable DC is indeed confirmed by numerical simulations. The PL with the perturbed refractive index can be fabricated via the femtosecond-laser direct-writing method [55] or optothermal nonlinearities [56] in a supporting silica media (the refractive index $n_0$ = 1.45). Each waveguide in a centered-square lattice with a period $a = 30\sqrt{2}$ $\mu$m only permits fundamental mode and is weakly coupled to the surrounding waveguides via a chain of waveguides. We choose the following parameters to implement the graphyne-like lattice: the perturbed refractive index is $\Delta n_0 = \Delta n_1 = 0.0016$, and the wavelength of laser is 1064 nm. The diameter is slightly tuned to keep DPs at the same level. Band structure has been calculated using mode analysis in software COMSOL 52a by finite element method [57]. After decreasing the refractive index $\Delta n_2$ to tune the hopping $t_2$, we attained the type-II, type-III and type-I DC, as shown in FIG. 3(b-g) and FIG. S2 (see supplementary materials). DPs have a slight shift along the $k_x$ direction, because the interaction between adjacent waveguide chains makes the hopping $t_1$ different.

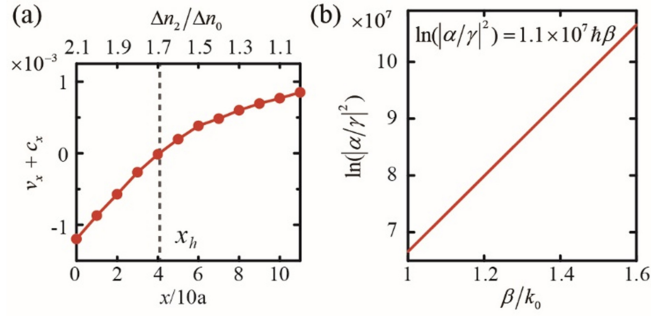

FIG. 4. (a) The calculated gradual evolution of group velocity for counter-propagating photon modes in an inhomogeneous PL with a horizon at $x_h$. Hawking temperature is ~0.14 $m$K in this design. (b) The associated logarithmic ratio of positive and negative norm amplitudes.

The optical BH analog in an inhomogeneous PL was finally accomplished by gradually decreasing the refractive index $\Delta n_2$ along the $x$ direction, $\Delta n_2(x) \propto -x\Delta n_0/100a$ [FIG. 4] (a white hole is also makeable just by reversing the BH, as shown in supplementary materials). The logarithmic ratio of positive and negative norm (amplitudes) maintains a linear relationship with (quasi)frequency, as shown in FIG. 4(b), where the slop is determined by Hawking temperature. Waveguide array can be readily fabricated by the femtosecond-laser direct-writing method [55], where refractive index of each waveguide is accurately tuned by adjusting writing speed or laser intensity. When the incoming modes are excited by an external light source in type-II region close to the horizon, the outgoing modes will be amplified by the transfer of incoming modes. Hawking radiation occurs during the filling of eigenmodes supported by isofrequency surface in type-II region. Hawking temperature will be obtained by detecting the emission intensity of the two modes. Type-I Dirac cones have a vanishing density of state (DOS) at the point-like isofrequency surface. While DOS becomes a parabolic peak for type-II Dirac cones at the conical-like isofrequency surface [36]. Consequently, spontaneous emission in proportion to DOS can be significantly enhanced in type-II region, when photons are supplied by an external light source, such as by embedding fluorescent molecules in the vicinity of the event horizon. We note that the band structure of PLs is special, which exhibits the diffraction of light in waveguide array using propagation constant $\beta$ at the same frequency. In fact, $c_{x(y)}$ and $v_x$ are not real group velocities but diffraction angles. Interestingly, the propagation direction along the waveguides acts as a temporal coordinate. The diffraction in the inhomogeneous PL is the same as light trajectory in spacetime. However, we need to correct Hawking temperature by multiplying the group velocity along the waveguide array. In doing so, we have created a curved spacetime permitting a Hawking temperature 0.14 $m$K.

The equations we employ here are general for other similar systems. For example, in photonic crystals whose bands manifest propagation of light with different frequencies in place of $\beta$, the same approach can be applied. Also, Hawking temperature can be increased by increasing the effective gravity at the horizon, which can be realized by enlarging the hopping or the evolutionary gradient of DC. While Hawking temperature is limited by diffraction angle in PL, it can reach above 1 K in photonic crystals.

In summary, we have demonstrated a novel optical analog of BH and Hawking radiation via topological phase transition. The gradual evolution from type-II to type-I region creates a BH spacetime with a robust long-lived event horizon at their interface. As an example, we propose an accomplishable laboratory BH analog in an inhomogeneous graphyne-like PL with Hawking temperature of 0.14 $m$K. We envision that this kind of optical BH analog can be generalized into other systems, such as photonic crystals [58, 59], metamaterials [60-62], plasmonic crystals [63, 64], transformation optics [65-67] and acoustics [68, 69]. BH laser [70] is also possible by designing with a pair of horizons of black and white holes as a resonator. Study of such BH analogs may reveal rich physics associated with their Hawking-like radiation and shed new lights to gravitational physics in association with topology.

**Acknowledgments**

Work at Wuhan was supported by the National Key Basic Research Program (Grant No. 2015CB932400) and the National Natural Science Foundation of China (Grant Nos. 11674255, 11674256). Work at Utah was supported by U.S. DOE-BES (Grant No. DE-FG02-04ER46148). M. K. acknowledges financial support from the Wuhan University graduate student overseas exchange program.